\documentclass{jkas}

\def\year{2014} % publication year
\def\month{May} % publication month
\def\volume{00} % journal volume
\def\issue{0} % journal issue
\def\beginpage{1} % first page of article
\def\endpage{99} % last page of article
\def\received{2015} % date paper was received by JKAS
\def\accepted{2015} % date of acceptance
\date{Received \received; accepted \accepted}

\title{SEJONG OPEN CLUSTER SURVEY (SOS) - V. THE ACTIVE STAR FORMING REGION SH 2-255 -- 257
}

\author[1]{Beomdu Lim\thanks{Korea Research Council of Fundamental Science \& Technology Research Fellow}}
\author[2]{Hwankyung Sung}
\author[2]{Hyeonoh Hur}
\author[1,3]{Byeong-Cheol Lee}
\author[4]{Michael S. Bessell}
\author[5]{Jinyoung S. Kim}
\author[6]{Kang Hwan Lee}
\author[1,3]{Byeong-Gon Park}
\author[1,3]{Gwanghui Jeong}

\affil[1]{Korea Astronomy and Space Science Institute, 776 Daedeokdae-ro, Yuseong-gu, Daejeon 305-348, Korea; \email{bdlim1210@kasi.re.kr}}
\affil[2]{Department of Astronomy and Space Science, Sejong University, 209 Neungdong-ro, Gwangjin-gu, Seoul 143-747, Korea; \email{sungh@sejong.ac.kr}}
\affil[3]{Astronomy and Space Science Major, University of Science and Technology, Gajeong-ro, Yuseong-gu, Daejeon 305-333, Korea}
\affil[4]{Research School of Astronomy and Astrophysics, Australian National University, MSO, Cotter Road, Weston, ACT 2611, Australia}
\affil[5]{Steward Observatory, University of Arizona, 933 N. Cherry Ave. Tucson, AZ 85721-0065, USA}
\affil[6]{Gwacheon National Science Museum, 110 Sanghabeol-ro, Gwacheon-si, Gyeonggi-do 427-060, Korea }
\def\corrauthor{Beomdu Lim}

\def\runningauthor{Lim et al.}

\def\runningtitle{THE HII REGIONS SH 2-255--257}

\def\keywords{
open clusters and associations: individual: Sh 2-255 -- 257 (IC 2162) -- dust, extinction -- stars: luminosity function, mass function.
}

\def\abstracttext{There is much observational evidence that active star formation is taking place in the H{\scriptsize \textsc{II}} 
regions Sh 2-255 -- 257. We present a photometric study of this star forming region (SFR) using imaging data obtained in passbands 
from the optical to the mid-infrared in order to study the star formation process. A total of 218 members were identified using various 
selection criteria based on their observational properties. The SFR is reddened by at least $E(B-V) = 0.8$ mag, and the reddening law 
toward the region is normal ($R_V = 3.1$). From the zero-age main sequence fitting method it is confirmed that the SFR is $2.1 \pm 0.3$ kpc 
from the Sun. The median age of the identified members is estimated to be about 1.3 Myr from comparison of the Hertzsprung-Russell 
diagram (HRD) with stellar evolutionary models. The initial mass function (IMF) is derived from the HRD and the near-infrared ($J, J-H$) 
color-magnitude diagram. The slope of the IMF is about $\Gamma = -1.6 \pm 0.1$, which is slightly steeper than that of the Salpeter/Kroupa 
IMF. It implies that low-mass star formation is dominant in the SFR. The sum of the masses of all the identified members provides the lower 
limit of the cluster mass ($169 M_{\odot}$). We also analyzed the spectral energy distribution (SED) of pre-main sequence stars using the 
SED fitting tool of Robitaille et al. and confirm that there is a significant discrepancy between stellar mass and age obtained from two different 
methods based on the SED fitting tool and the HRD. }

\begin{document}
\jkashead %% set title, authors, abstract, etc.

%%%%%%%%%%%%%%%%%%%%%%%%%%%%%%%%%%%%%%%%%%%%%%%%%%%%%%%%%%%%%%%%%%%%%
%%% BEGIN MAIN TEXT HERE %%%%%%%%%%%%%%%%%%%%%%%%%%%%%%%%%%%%%%%%%%%%
%%%%%%%%%%%%%%%%%%%%%%%%%%%%%%%%%%%%%%%%%%%%%%%%%%%%%%%%%%%%%%%%%%%%%

\section{INTRODUCTION}
Young open clusters provide useful test beds for the study of star formation processes because about 80 -- 90\% of young stars are 
found in embedded clusters with more than 100 members \citep{LL03,PCA03}. Furthermore, the fundamental parameters of 
clusters such as reddening, distance, and age can be properly constrained. These advantages allow us to derive a more reliable stellar 
initial mass function (IMF) with which to investigate star formation processes. 

Sh 2-254 -- 258 is a famous star forming region (SFR) in the Gem OB1 association \citep{Sh59}. The main ionizing sources 
of the H{\scriptsize \textsc{II}} regions are known to be one late-O and four early-B-type stars \citep{CAH08}. A number 
of previous works found a few maser sources \citep{T71,LB73,GMS07} as well as various molecular lines \citep{MPTZ74,
BPvB75,EBB77,BPV09,ZLS12} in the region. Several sub-structures such as clumps and cores were also reported 
from infrared (IR), sub-millimeter, millimeter, and centimeter observations \citep{BBW79,JDDH84,HPF97,MBH05,ZLS12}. These 
observational properties commonly indicate that active star formation is in progress.   

\begin{figure*}[t!]
\centering
\includegraphics[angle=0,width=70mm]{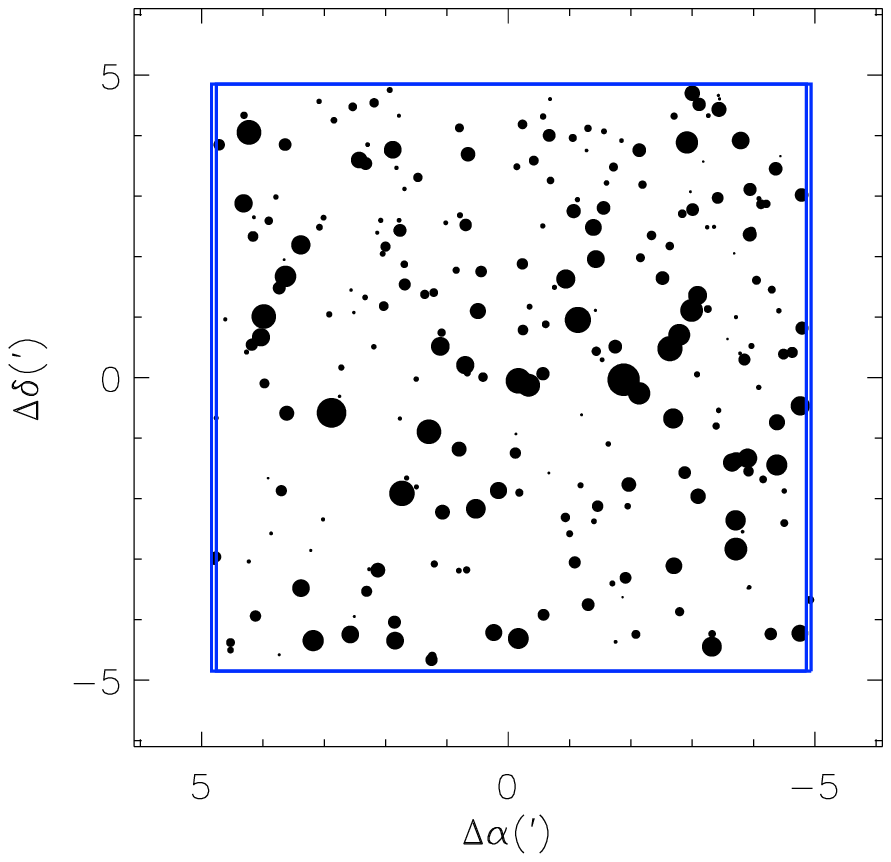}\includegraphics[angle=0,width=55mm]{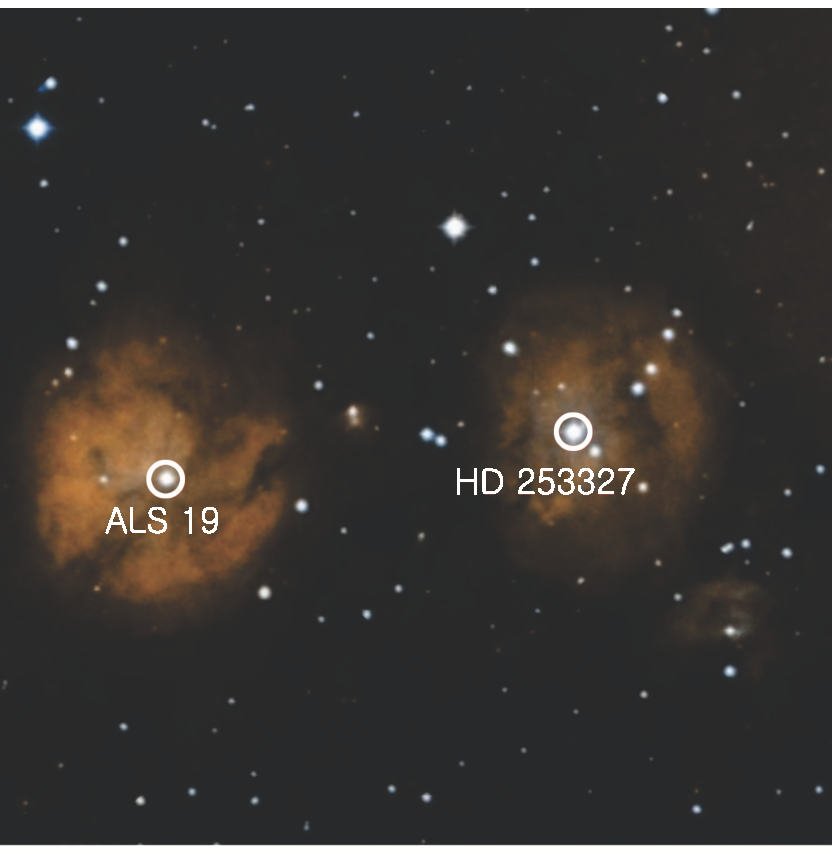}
\includegraphics[angle=0,width=70mm]{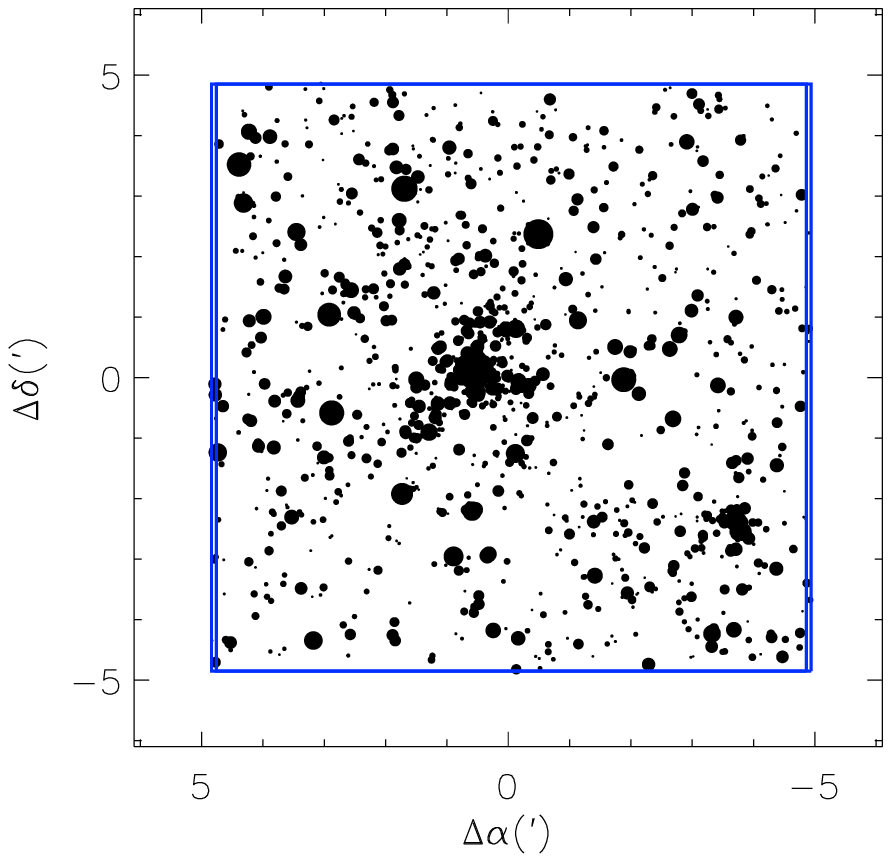}\includegraphics[angle=0,width=55mm]{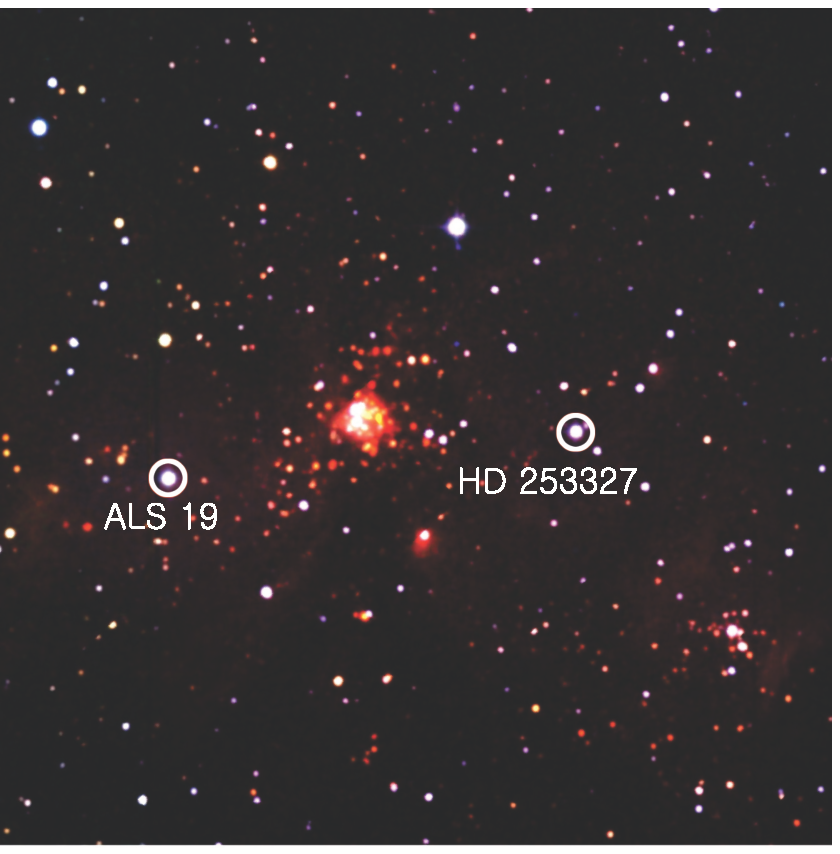}
\caption{Finder chart and color composite image of the observed region in optical passbands (upper) and 
near-infrared passbands (lower). Stars brighter than $V = 18$ mag and $K_S = 16$ mag are plotted in left-hand panels, respectively. 
The size of the circles is proportional to the brightness of individual stars. The position of stars are relative to $\alpha = 06^{\mathrm{h}} 
\ 12^{\mathrm{m}} \ 52^{\mathrm{s}}.1, \ \delta = +17^{\circ} \ 59' \ 16''.4$. Squares outlined by blue solid lines represent the field of view of 
the Mont4K CCD camera. The color composite images were obtained from the Digital Sky Survey-2 and Two Micron All Sky Survey. 
The position of two ionizing sources is marked by open circles in each image. }\label{fig1}
\vspace{5mm} %% add extra space ONLY when figures/tables are "colliding"!
\end{figure*}

\citet{CAH08} took a census of young stellar objects (YSOs) using extensive near- to mid-IR imaging data. Most of the 
young stars ($\sim 80$\%) were found in several embedded sub-clusters. \citet{MPZ11} continued their search for low-mass 
YSOs in the quiescent phase, with a deep {\it Chandra} X-ray observation, and found that the total number of YSOs was consistent with 
that expected from the Kroupa IMF \citep{K01} scaled to the number of ionizing sources. On the other hand, sequential 
star formation scenarios within the SFR have been proposed \citep{HPF97,MPL07,CAH08,BPV09,WBB11,MPZ11}. Circumstantial 
evidence, such as the age difference among H{\scriptsize \textsc{II}} regions, and the number ratio of YSOs at different evolutionary 
stages, indicate that star formation activity propagated from Sh 2-255 and 257 into the molecular clouds behind the H{\scriptsize \textsc{II}} 
bubbles. Hence, this SFR is one of the more interesting sites in which to study star and cluster formation processes.

The present work on the embedded young open clusters in the H{\scriptsize \textsc{II}} regions Sh 2-255 -- 257 (hereinafter IC 2162) 
is the sixth paper of the Sejong Open cluster Survey (SOS) project. \citet[hereinafter Paper 0]{SLB13} presented the 
overview of the SOS project. Comprehensive studies of several open clusters IC 1848, NGC 1624, NGC 1893, NGC 1931, 
and NGC 2353 were carried out as part of the project \citep{LSKI11,LSK14a,LSK14b,LSB15}. In this current work, we 
revise the fundamental parameters of the SFR in a homogeneous manner, and constrain the IMF to study 
star formation processes. The observational data we used are described in Section 2. In Section 3, we present several fundamental 
parameters of IC 2162 obtained from photometric diagrams and discuss the reddening law toward the SFR. 
The IMF is derived in Section 4, and the spectral energy distribution (SED) of pre-main sequence (PMS) members is investigated 
in Section 5. Finally, the comprehensive results from this study are summarized in Section 6.

\section{OBSERVATIONAL DATA}

\subsection{Optical Imaging Data}
The observations of IC 2162 were made on 2013 February 5, using the Kuiper 61" telescope (f/13.5) of Steward Observatory 
on Mt. Bigelow in Arizona, USA. Images were taken with the Mont4K CCD camera and 5 filters (Bessell $U$, Harris $BV$, 
Arizona $I$, and H$\alpha$) in a $3 \times 3$ binning mode. The field of view (FOV) is about $9.^{\prime}7 \times 9.^{\prime}7$. 
The target images comprise 12 frames that were taken in two sets of exposure times for each band (5 and 180 s $\times$ 2 
in $I$, 5 and 180 s $\times 2$ in $V$, 7 and 300 s in $B$, 30 and 600 s in $U$, and 30 and 600 s in H$\alpha$). We also 
observed several equatorial standard stars \citep{MML91} at air masses of 1.2 -- 2 on the same night in order to transform 
the instrumental magnitudes to the standard magnitude and colors. Additional standard stars with extremely blue and red colors 
in the Landolt standard star field Rubin 149 \citep{L92} were observed to determine the secondary extinction coefficients.  

\begin{table}[t]
\caption{Atmospheric Extinction Coefficients and Photometric Zero points \label{tab1}}
\centering
\begin{tabular}{lcccc}
\toprule
Filter & $k_1$ & $k_2$ & $\zeta$ (mag) \\ 
\midrule
$I$ & $0.045 \pm 0.008$ & - & $22.170 \pm 0.009$ \\  
$V$ & $0.120 \pm 0.008$ & - & $23.560 \pm 0.007$ \\
$B$ & $0.232 \pm 0.008$ & $0.023 \pm 0.002$ & $23.548 \pm 0.006$ \\  
$U$ & $0.444 \pm 0.018$ & $0.031 \pm 0.005$ & $22.069 \pm 0.008$ \\
H$\alpha$ & 0.085 &  - & 19.565\\
\bottomrule
\end{tabular}
\end{table}

All the pre-processing to remove the instrumental signals were carried out using the \textsc{IRAF}\footnote{Image Reduction 
and Analysis Facility is developed and distributed by the National Optical Astronomy Observatories, which is operated 
by the Association of Universities for Research in Astronomy under cooperative agreement with the National Science 
Foundation.}/\textsc{CCDRED} packages. Simple aperture photometry was performed for the standard 
stars with an aperture size of $14.^{\prime \prime}0$ (16.3 pixels). The primary and secondary atmospheric extinction 
coefficients were determined from the photometric data of the standard stars using a weighted least-square method. 
We present the coefficients and photometric zero points in Table~\ref{tab1}. Point spread function (PSF) photometry of 
stars in the target images was performed with a small fitting radius of one full width at half-maximum ($\leq 1.^{\prime \prime}0$) 
using \textsc{IRAF/DAOPHOT}. Aperture photometry of bright, isolated stars with a photometric error smaller 
than 0.01 mag in individual target images was obtained to correct for the aperture difference. The instrumental magnitudes 
of stars in the target images were transformed to the standard magnitude and colors using the transformation relations 
as described in Appendix of \citet{LSB15}. The finder chart for the stars brighter than $V = 18$ mag is 
shown in the upper left-hand panel of Figure~\ref{fig1}. 

A total of 811 stars were detected from optical photometry. The completeness of our photometry was assessed from 
the luminosity function of all observed stars. The luminosity function exhibits a single linear slope in the 
magnitude range of $V = 13 - 19$ mag. If we assume that the linear slope is applicable down to the faint stars, the 
turn-over magnitude gives the completeness limit. As a result, our photometry seems to be about 90\% complete down 
to $V = 19.3$ mag. 

\begin{figure*}[t!]
\centering
\includegraphics[angle=0,width=150mm]{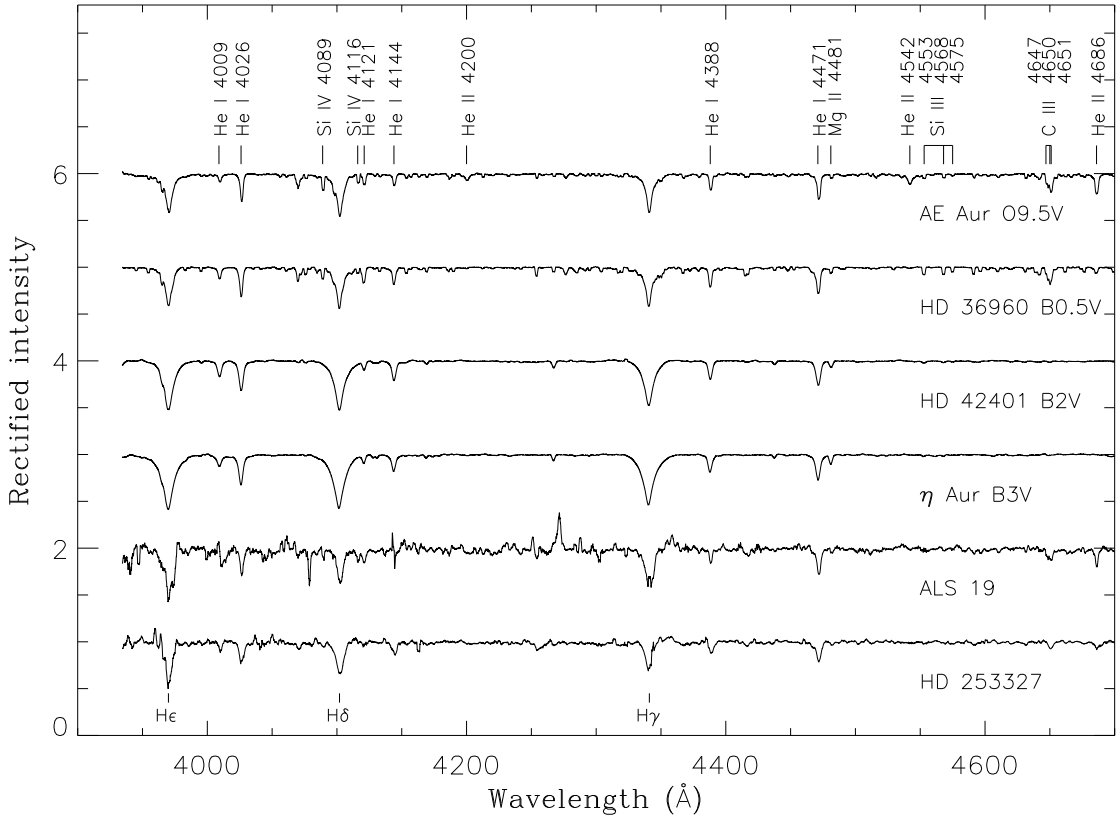}
\caption{Optical spectra of four standard stars and two ionizing sources (from top to bottom) in the observed field of view. 
The object name is denoted below each spectrum. Main spectral lines used in spectral classification are identified at the top of the 
figure.  }\label{fig2}
\vspace{5mm} %% add extra space ONLY when figures/tables are "colliding"!
\end{figure*}

\subsection{Optical Spectroscopic Data}
The optical spectra of two main ionizing sources (ALS 19 and HD 253327) were obtained on 2015 March 10 with the 
fiber-fed echelle spectrograph BOES (Bohyunsan Observatory Echelle Spectrograph -- \citealt{KHV07}) attached to the 1.8 m telescope 
at Bohyunsan Optical Astronomy Observatory in Korea. A single frame for each target was taken with a 300 $\mu m$ 
fiber ($R = 30,000$), and the exposure time was 3600 seconds. The 3 $\times$ 3 binning mode allowed us to improve 
the signal-to-noise ratio of the spectra. For the wavelength calibration, spectra of a ThAr lamp were also acquired 
on the same night. 

Pre-processing and extraction of spectra were made with the \textsc{IRAF}/\textsc{ECHELLE} package. 
A sigma clipping method was used to minimize the influence of cosmic rays on the individual frames in a given order. We 
normalized the spectra using the best solution found from a cubic spline interpolation and finally smoothed them by 
a box size of 33. The spectra of ALS 19 and HD 253327 are shown in Figure~\ref{fig2}. For comparison, the spectra 
of standard stars [AE Aur (O9.5V),  HD 36960 (B0.5V), HD 42401 (B2V), and $\eta$ Aur (B3V) --  \citealt{W71,SMW11}] 
observed with the same instrument on 2014 October 29 are also plotted in the same figure.

\subsection{Archival Infrared Data}
We transformed the CCD coordinates ($x_{\mathrm{CCD}}, y_{\mathrm{CCD}}$) of the optical photometric data 
into celestial coordinates ($\Delta \alpha, \Delta \delta$) using the Two Micron All Sky Survey catalogue 
(2MASS; \citealt{2mass}). Optical counterparts of near-IR sources in the 2MASS catalogue were searched 
for with a matching radius of $1^{\prime \prime}$. A total of 361 optical counterparts were found.

\citet{CAH08} has made an extensive IR imaging survey across the entire molecular complex 
incubating the H{\scriptsize \textsc{II}} regions Sh 2-254 -- 258. This survey covers an area of $25^{\prime} \times 
20^{\prime}$. Their catalogue includes the near-IR $JHK_S$ and {\it Spitzer} InfraRed Array Camera (IRAC) 4-band 
photometry of 26,821 sources. The near-IR photometry in the IR source catalogue is reasonably tied to the 2MASS 
photometric system within 0.03 mag. Only stars within the FOV of the optical imaging observations
were used in our analysis. A total of 3,426 sources were found within our FOV ($\sim 9.^{\prime}7 \times 9.^{\prime}7$), 
of which 792 sources have optical counterparts within a matching radius of $1^{\prime \prime}$. We present the finder chart of these 
stars in the lower left panel of Figure~\ref{fig1}.

A post-BCD (basic calibrated and mosaiced) image of the {\it Spitzer} Multiband Imaging Photometer (MIPS) 
24 $\mu$m image was taken from the data archive of the {\it Spitzer} Science Center (ObsID: 40005, PI G. Fazio). We carried 
out PSF photometry for stars in the image using the \textsc{IRAF/DAOPHOT} with a fit radius of 2.4 pixel and a sky annulus of 
20$^{\prime\prime}$ -- 32$^{\prime\prime}$ (see \citealt{SSB09}). The photometric zero point of 11.76 mag was calculated using the pixel 
scale and the flux of a zeroth magnitude star as described in MIPS Handbook. Within our FOV, a total of 207 sources 
were detected, of which 13 and 30 sources have counterparts in the optical and IR catalogues, respectively.

\section{FUNDAMENTAL PARAMETERS}
\begin{figure*}[t!]
\centering
\includegraphics[angle=0,width=150mm]{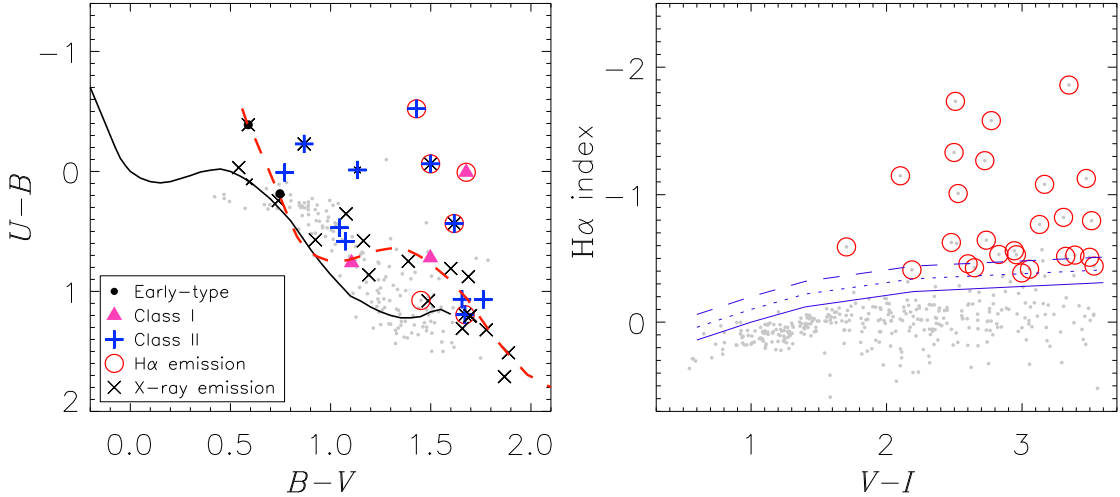}
\caption{Color-color diagrams of stars in IC 2162. The small dots (grey) represent all the stars. Other symbols 
denote early-type members (black bold dots), Class I (magenta triangles), Class II (blue pluses), X-ray emission stars (large crosses),  
X-ray emission candidates (small crosses), and H$\alpha$ emission stars or candidates (red open circles), respectively. The solid 
line (black) in the left-hand panel exhibits the intrinsic color-color relation of \citet{SLB13}, and its reddened relation [$E(B-V) = 0.88$ 
mag] is shown by a dashed line (red). The solid line in the right-hand panel represents the empirical photospheric level of unreddened 
main sequence stars, while the dashed and dotted lines are the lower limit of H$\alpha$ emission stars and H$\alpha$ emission candidates, 
respectively.}\label{fig3}
\vspace{5mm} %% add extra space ONLY when figures/tables are "colliding"!
\end{figure*}

\begin{figure*}[t!]
\centering
\includegraphics[angle=0,width=150mm]{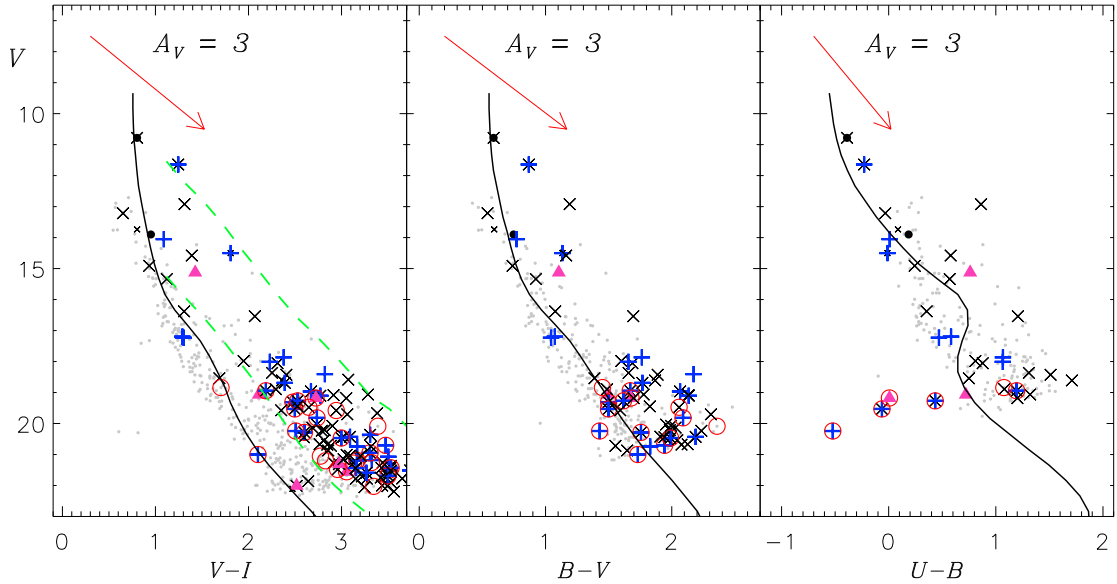}
\caption{Color-magnitude diagrams in the optical passbands. Left-hand panels: $V-I$ versus $V$ diagram. The location of 
pre-main sequence stars is confined between the green dashed lines. Middle panels: $B-V$ versus $V$ diagram. Right-hand 
panels: $U-B$ versus $V$ diagram. The solid lines represent the reddened zero-age main sequence relation of \citet{SLB13}. The 
arrow denotes the reddening vector corresponding to $A_V = 3$ mag. The other symbols are the same as Figure~\ref{fig3}.}\label{fig4}
\vspace{5mm} %% add extra space ONLY when figures/tables are "colliding"!
\end{figure*}

As seen in Figure~\ref{fig1}, the finder charts and color composite images in the optical and near-IR passbands 
exhibit completely different stellar distributions. This implies that the majority of young stars are embedded behind the H{\scriptsize \textsc{II}} 
regions. Because only about a quarter of the stars were detected in the optical passbands, the canonical analysis based on the optical 
photometric diagrams is limited. For this reason, the IR photometry of \citet{CAH08}, which is less sensitive to the effect of 
extinction, is a powerful tool to probe embedded populations. However, the several visible stars are still very helpful to 
determine fundamental parameters such as reddening and distance. In this section, we describe the identification of the
main ionizing sources, membership selection criteria, and the determination of reddening, distance, and age based on 
the optical spectra and photometric diagrams as presented in Figure~\ref{fig2} -- \ref{fig4}.

\subsection{Spectral Types of Two Ionizing Sources}
The influence of high-mass stars on the surrounding environment involves destructive and constructive processes. The 
strong stellar wind and radiation pressure of high-mass stars can disperse their natal clouds, and thereby terminate 
star formation. On the other hand, H{\scriptsize \textsc{II}} bubbles created by these stars can accumulate and compress 
material as they expand into the molecular clouds. The condensed material can then form a new generation of 
stars \citep{EL77}. The high-mass stars can also drive the formation of the second generation of stars radiatively in 
pre-existing clumps \citep{LL94}. Therefore, the identification of the ionizing sources is essential for studying such 
feedback of high-mass stars.

The brightest stars ALS 19 and HD 253327 (ALS 18) are known to be the main ionizing sources of IC 2162. In order to 
classify the spectral type of these stars, we adopted the O and B-type star classification scheme of \citet{WF90,SMW11}. 
The spectra of the stars in Figure~\ref{fig2} contain several emission-like features that are the residuals of cosmic rays. 
In the case of ALS 19, He {\scriptsize \textsc{II}} $\lambda$4200 and $\lambda$4542 are invisible in the spectrum, while 
He {\scriptsize \textsc{II}} $\lambda$4686 absorption is clearly seen. The spectral type of this star is likely to be B0V; however, it is also 
possible that ALS 19 is a late-O-type star (O9.5V or O9.7V) given the strength of He {\scriptsize \textsc{II}} $\lambda$4686 
and the line ratio between Si {\scriptsize \textsc{IV}} $\lambda$4116 and He {\scriptsize \textsc{I}} $\lambda$4121. Since 
the star was classified as Class II \citep{CAH08}, the spectrum of this young star shows a mixture of late-O and 
early-B-type star characteristics. 

He {\scriptsize \textsc{II}} $\lambda$4200 and $\lambda$4542 are also absent in the spectrum of HD 253327. Si {\scriptsize \textsc{III}} 
$\lambda$4552 is invisible, while a weak He {\scriptsize \textsc{II}} 4686 absorption is authentically seen. Hence, the spectral type of 
HD 253327 is likely to be B0V. The spectrum of this star is, indeed, similar to that of ALS 19. Our spectral classification is in a good 
agreement with that of previous studies \citep{GGR73,CAH08}.

\subsection{Membership Selection}
Early-type main sequence (MS) stars (putatively B-type stars) can be selected from the optical photometric diagrams 
(Figure~\ref{fig3} and ~\ref{fig4}) as they are bright in V and very blue in $U-B$. In addition, the reddening 
and distance of the individual stars can be reliably determined because the intrinsic colors and absolute magnitude of such 
stars have been well calibrated in the optical passbands. Probable early-type members are firstly selected from magnitude and 
color cuts as $V \leq 15$ mag, $0.5 \leq B-V \leq 0.9$, $-0.6 \leq U-B \leq 0.5$, and $Q^{\prime} \leq -0.3$, where $Q^{\prime} 
\equiv (U-B) - 0.72(B-V) - 0.025E(B-V)^2$ (Paper 0). We then removed several foreground late-type stars (probably F- or G-type) 
restricting the reddening range to $E(B-V) > 0.8$ mag and color excess ratios. In addition, a few stars identified as YSOs 
by \citet{CAH08} were also excluded from the MS member list. Only two stars were finally selected as the early-type 
MS members of IC 2162.

We utilized H$\alpha$ photometry as a criterion to identify PMS stars in the SFR. A series of studies 
demonstrated that H$\alpha$ photometry can effectively detect a number of low-mass PMS stars at the T Tauri stage in 
young open clusters \citep{SBL97,SBL98,SCB00,PSBK00,PS02,SBC04,SB04,SBC08,SSB13,LSK14a,LSK14b,LSB15}. 
In order to detect objects with an H$\alpha$ emission line, the H$\alpha$ index [$\equiv$ H$\alpha - (V + I)/2$] is used as the detection criterion \citep{SCB00}. 
As shown in the right-hand panels of Figure~\ref{fig3}, stars with an H$\alpha$ index smaller than the empirical photospheric level 
(solid line) of normal MS stars by $-0.2$ (dashed line), or $-0.1$ mag (dotted line), was selected as H$\alpha$ emission stars and 
candidates, respectively. We found 21 H$\alpha$ emission stars and six candidates. However, the H$\alpha$ emission star ID 659 ($V = 18.84$, 
$V-I = 1.70$, $B-V = 1.45$, and $U-B =  1.07$) is likely an active late-type star in the field because its colors are similar to those of other field 
stars. A total of 26 H$\alpha$ emission stars and candidates were selected as PMS members of IC 2162.

Excess emission at IR wavelengths, particularly the mid-IR, is a useful membership 
selection criterion because a large fraction of PMS stars in young open clusters ($\leq$ 3 Myr) have been found to have warm 
circumstellar disks or envelopes \citep{LMH00,SSB09,BNM13}. \citet{CAH08} identified 252 YSOs (87 Class I and 165 Class II) in the 
H{\scriptsize \textsc{II}} regions Sh 2-254 -- 258 using {\it Spitzer}/IRAC images. We used their YSO list and found 
optical counterparts for 64 YSOs (11 Class I and 53 Class II) within our FOV. However, only 41 IR sources 
(6 Class I and 35 Class II) were detected in the $V$ band.

PMS stars are also known as X-ray emitting objects \citep{FMS99,SBC04,CMP12,HSB12,SSB13,LSK14b,HPS15}. \citet{MPZ11} 
made deep X-ray observations of these SFRs down to 0.5 $M_{\odot}$ with the {\it Chandra} X-ray observatory. The observations covered a $17^{\prime} 
\times 17^{\prime}$ field, and detected a total of 364 X-ray sources. We used their X-ray source list to select 
X-ray emitting PMS members. The optical counterparts of the X-ray sources and candidates were searched for with matching 
radii of $1.^{\prime \prime}0$ and $1.^{\prime \prime}5$, respectively. We confirmed that 86 X-ray sources and three candidates were detected in the $V$ band. 
Among them 73 X-ray sources and one candidate are associated with PMS stars, and the early-type MS star HD 253327 
also turns out to be an X-ray source. The other 14 sources and candidates seem to be X-ray active field stars from their colors. 

A total of 102 members were identified in the optical passbands. In addition, members of IC 2162 were independently selected from 
the IR source catalogue of \citet{CAH08} in the same way. We found 216 members observed in the $J$ and $H$ bands (2 early-type 
MS stars, 20 Class I, 61 Class II, 158 X-ray sources, 7 X-ray candidates, 19 H$\alpha$ emission stars, and 6 H$\alpha$ candidates). 
The membership list from the IR source catalogue (216 stars) was compared with that from the optical data (102 stars). As a result, 
100 stars were found in both lists, however the other two members were observed only in either the $J$ or $H$ band. These 
membership lists were merged into a membership catalogue. The total number of members identified in this work is 218. 

\citet{MPZ11} estimated about 58 field interlopers scaling the number of contaminants in the FOV of the {\it Chandra} Carina 
Complex Project \citep{BGP11} to that in the FOV of $17^{\prime} \times 17^{\prime}$. If we assume that the surface density 
of the field interlopers is uniform across these SFRs, the number of the field interlopers in the FOV of $\sim 9.7^{\prime} \times 
9.7^{\prime}$ is about 18. According to the X-ray source classification toward the Carina region \citep{GBF11,BGP11}, 
extragalactic sources are so faint that the contribution of these sources should be negligible in this study. Probable interlopers 
in our FOV may therefore be foreground and background stars. However, as the stellar density in the direction of IC 2162 is much 
lower than that of the Carina region located toward the tangential direction of the Sagittarius spiral arm, the 
expected number of field interlopers with X-ray emission in our FOV may be much smaller than 18. As we identified 
14 X-ray sources and candidates as field interlopers, the number of field interlopers in our member catalogue may be less than four.
 
\begin{figure*}[t!]
\centering
\includegraphics[angle=0,width=150mm]{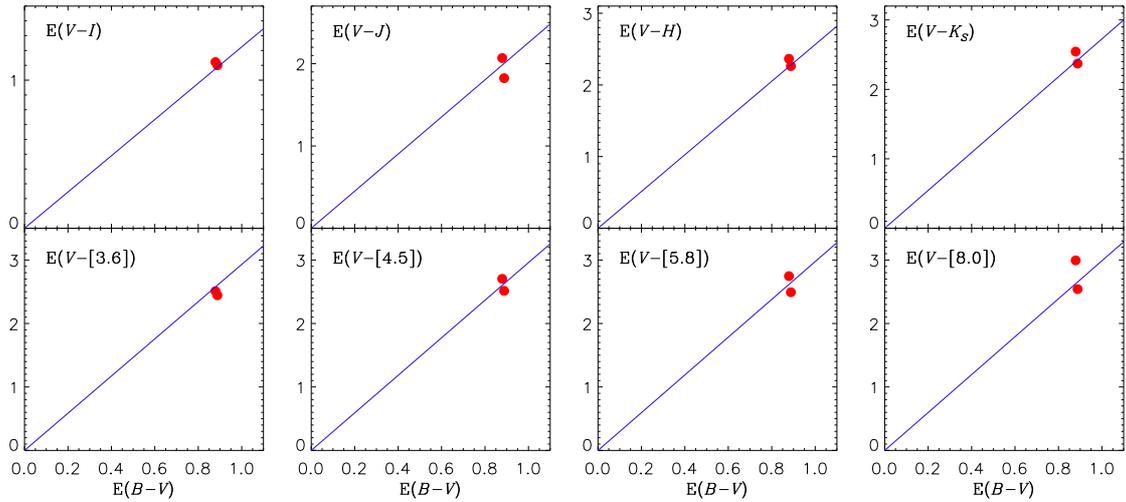}
\caption{Color excess ratios obtained from the early-type main sequence members. The solid line corresponds to $R_V = 3.1$. 
The color excess ratios from optical to mid-infrared data suggest that the reddening law toward IC 2162 
is normal.}\label{fig5}
\vspace{5mm} %% add extra space ONLY when figures/tables are "colliding"!
\end{figure*}

\subsection{Interstellar Extinction and the Reddening Law}

Light from stars is obscured by the interstellar material distributed along the line of sight. Therefore, the 
effect of extinction on the measured flux or magnitude should be corrected properly. The reddening of young open clusters 
can be estimated by comparing the observed colors of early-type stars with their intrinsic colors in the ($U-B$, $B-V$) two color 
diagram along the reddening slope (Paper 0). The reddening determined from two early-type MS members is about $E(B-V) = 0.88$ 
mag. In addition, the spectral type of two early-type PMS members ALS 19 (B0V) and 2MASS J06123651+1756548 (B0.9V) 
is available from this work and \citet{CAH08}. The reddening of these stars obtained from the spectral type-color relation 
(Table 5 in Paper 0) is about $E(B-V) = 1.41$ and 1.17 mag, respectively. The result is consistent with that of previous studies, 
e.g. $E(B-V) =$ 0.88 -- 1.16 mag \citep{PH76}, 0.64 -- 1.47 mag \citep{MJF79}, and 0.82 -- 1.20 mag \citep{HM90}. More 
severe differential reddening is expected across IC 2162 because a large number of stars are embedded in the molecular 
clouds. 

The ratio of total-to-selective extinction ($R_V$) is an essential diagnostic tool to investigate the extinction law toward SFRs or young 
open clusters. The general interstellar medium (ISM) in the solar neighbourhood is known to have, on average, $R_V  = 3.0$ -- 3.1 
\citep{FM07,LSKI11}. On the other hand, $R_V$ may be larger than the normal value in some dusty SFRs \citep{G10}. 
The extinction law depends on the size distribution of dust grains \citep{D03}. A large $R_V$ implies that the size of dust grains is, 
on average, larger than that found in the general ISM.

We investigated the various color excess ratios of the early-type MS members to study the reddening law toward IC 2162. $R_V$ 
can be determined from the color excess ratios between two different colors \citep{GV89,SSB13}. The color excess $E(V-\lambda)$ 
(where $\lambda = I, J, H, K_S$, [3.6], [4.5], [5.8], and [8.0]) can be computed from the intrinsic color relations of Paper 0 and 
Sung et al. (in preparation). Figure~\ref{fig5} displays the color excess ratios of the early-type MS members. 

The color excess ratios at different wavelengths are reasonably matched with the slope corresponding to $R_V = 3.1$ (solid lines in the figure). 
This result is acceptable given the $R_V$ variation with the Galactic longitude \citep{W77,SB14}. The normal reddening law implies that the 
dust evolution in the front side of the region had already progressed. However, this result may not represent the reddening law of the 
embedded cluster.

\begin{figure*}[t!]
\centering
\includegraphics[angle=0,width=170mm]{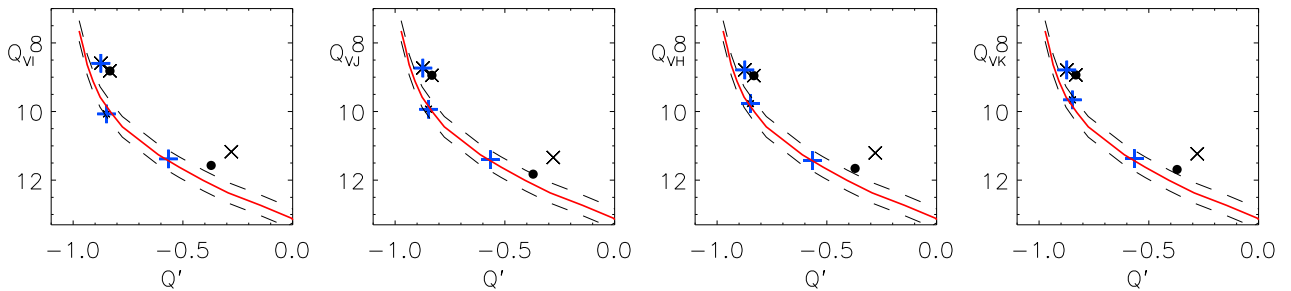}
\caption{Zero-age main sequence (ZAMS) fitting to the bright members in the $Q_{V \lambda}$-$Q^{\prime}$ diagrams. The 
ZAMS relations of \citet{SLB13} are fitted to the lower ridge line of the members. The solid line (red) represents the 
adopted distance modulus of 11.6 mag, equivalent to 2.1 kpc. The dashed lines indicate a 0.3 mag error in the ZAMS fitting. 
The other symbols are the same as Figure~\ref{fig3}.}\label{fig6}
\vspace{5mm} %% add extra space ONLY when figures/tables are "colliding"!
\end{figure*}

\subsection{Distance}
The distance to an object is fundamental in determining its physical quantities. We determined the 
distance of IC 2162 using the zero-age main sequence (ZAMS) fitting method. The ZAMS relations and reddening-independent indices 
introduced in Paper 0 are adopted in the present work. Our ZAMS fitting procedure is based on $UBVIJHK_S$ multicolor photometry, 
and therefore the distance can be determined consistently in the optical and near-IR passbands. 

Figure~\ref{fig6} shows $Q_{V \lambda}$-$Q^{\prime}$ diagrams of the bright members ($J < 12.5$). Since the luminosity of stars 
can be affected by stellar evolution and binary effects, the ZAMS relations should be fitted to the lower ridge line of the MS 
in the $Q_{V \lambda}$-$Q^{\prime}$ planes as shown in the figure. We adjusted the ZAMS relations above and below the faint members 
at a given $Q^{\prime}$ index and adopted a distance modulus of 11.6 mag. The lower ridge line could be confined between the ZAMS 
relations shifted from the adopted value by $\pm 0.3$ mag, and therefore the upper and lower envelopes are the uncertainty in the distance. 
Our result ($2.1 \pm 0.3$ kpc) is in reasonable agreement with that of previous studies within the uncertainties, e.g. 1.9 kpc 
\citep{CLH87,AH81}, 2.4 kpc \citep{HM90}, and 2.5 kpc \citep{PH76,MJF79,RAG07}.

\subsection{Age}
The Hertzsprung-Russell diagram (HRD) provides a comprehensive view of the evolutionary status of stellar system. The reddening of the individual 
members was corrected for using the weighted mean reddening of four early-type MS and PMS members, where the weight is exponentially 
decreased as the distance between the early-type stars and a given member increases. We converted the reddening-corrected color-magnitude 
diagrams (CMDs) in the optical passbands to the HRD using several relations (\citealt{B95,BCP98}; Paper 0). The effective temperature 
$T_{\mathrm{eff}}$ of a star was basically estimated from the color-$T_{\mathrm{eff}}$ relations, and the spectral type-$T_{\mathrm{eff}}$ relation 
was also used for the three early-type members ALS 19, HD 253327, and 2MASS J06123651+1756548. The weighted mean value of the temperatures was adopted 
as the $T_{\mathrm{eff}}$ of four early-type MS and PMS members. Only the $V-I$ versus $T_{\mathrm{eff}}$ relation 
\citep{B95,BCP98} was used for the temperature scale of the PMS members to avoid the effect of excess emission due to accretion activities. 
Bolometric correction values were estimated from the $T_{\mathrm{eff}}$ of the individual members using Table 5 of Paper 0. 

\begin{figure}[t!]
\centering
\includegraphics[angle=0,width=80mm]{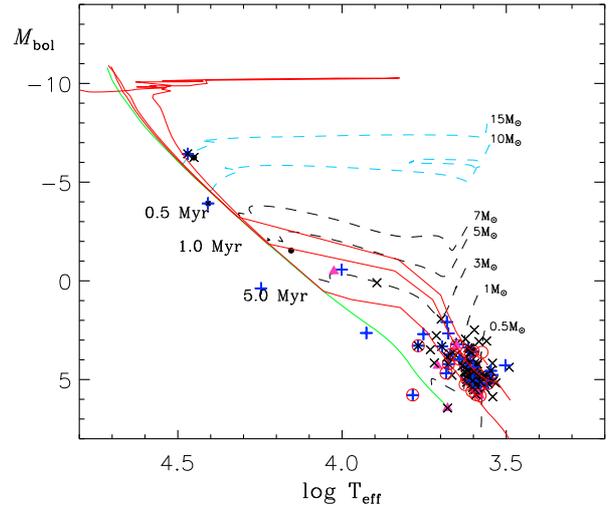}
\caption{Hertzsprung-Russell diagram of IC 2162. A few isochrones (red solid lines) for the age of 0.5, 1, and 5 Myr are superimposed 
on the diagram with the evolutionary tracks of several initial masses \citep{SDF00,EGE12}, where the solar metallicity is assumed. The 
other symbols are the same as Figure~\ref{fig3}}\label{fig7}
\vspace{5mm} %% add extra space ONLY when figures/tables are "colliding"!
\end{figure}

We present the HRD of IC 2162 in Figure~\ref{fig7}. The mass of the main ionizing sources in Sh 2-255 and 257 is larger than 
$10 M_{\odot}$. It is difficult to estimate the age from the MS turn-off because the stars seem to be still at the MS or PMS stage. On 
the other hand, the majority of the PMS members are evolving along Hayashi tracks, while only several members are approaching 
the ZAMS along Henyey tracks. There are five PMS members near or below the ZAMS. These stars may have edge-on disks 
\citep{SBL97,SB10}. The upper mass range of the PMS members appears to be as large as $15M_{\odot}$, and the PMS lifetime of high-mass stars 
is very short. These facts imply that IC 2162 is very young. Using the evolutionary models for PMS stars \citep[$Z=Z_{\odot}$ with 
convective overshooting]{SDF00}, we estimated the ages of the PMS members and found a median age of 1.3 Myr with an age spread of 3.3 Myr. 
The age distribution is very similar to that obtained from the evolutionary models for half solar metallicity ($Z=0.01$). The age 
of IC 2162 is probably about 1 Myr. However, as the majority of PMS  members are still deeply embedded in the molecular clouds, the age of 
these stars may be much younger.

\section{TOTAL MASS AND THE INITIAL MASS FUNCTION}

The masses of optically visible members are estimated by comparing their $T_{\mathrm{eff}}$ and $M_{\mathrm{bol}}$ with those 
of the evolutionary tracks in the HRD. The evolutionary tracks of \citet{EGE12} were used for the early-type MS and PMS members, 
while those of \citet{SDF00} were adopted for the low-mass PMS members. The solar metallicity was assumed for both models. 
These theoretical evolutionary masses can be used as a good mass scale reference for those inferred from the near-IR data.

\begin{figure}[t!]
\centering
\includegraphics[angle=0,width=75mm]{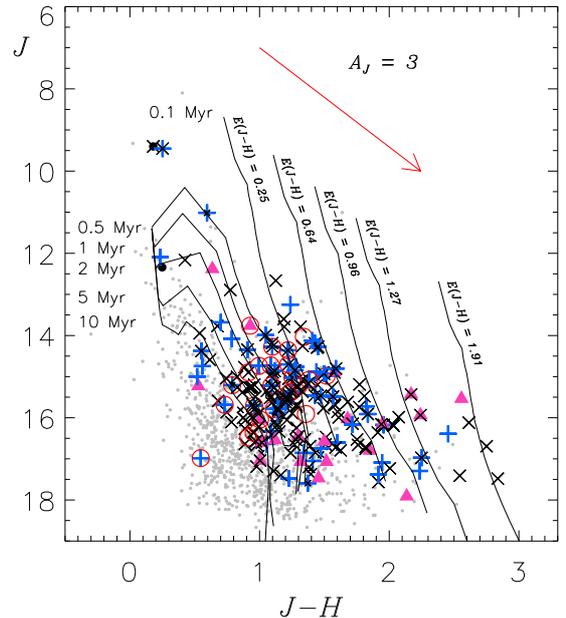}
\caption{Near-infrared color-magnitude diagram. The solid lines represent pre-main sequence star isochrones with different ages 
(0.1, 0.5, 1, 2, 5, and 10 Myr) from \citet{SDF00}, where the model parameters are the same as seen in Figure~\ref{fig7}. The isochrones are 
reddened by $E(J-H) = 0.25, 0.64, 0.96, 1.27$, and 1.91 mag, respectively, after correction for the distance modulus of 11.6 mag. The 
arrow denotes the reddening vector corresponding to $A_J = 3$ mag. The other symbols are the same as Figure~\ref{fig3}.
}\label{fig8}
\vspace{5mm} %% add extra space ONLY when figures/tables are "colliding"!
\end{figure}

The masses of the members selected from the IR source catalogue were inferred from the $(J, J-H)$ diagram because the $J$ magnitude is 
less affected by excess emission from a disk or envelope compared to the $K$ magnitude \citep{KNS07}. Figure~\ref{fig8} shows the 
near-IR CMD of IC 2162. The PMS members have a wide color range from  $J-H = 0.5$ to 2.9. A few factors such as high differential 
reddening, excess emission from disks and envelopes in the $H$ band, age spread, and photometric errors for faint stars may be responsible 
for such a large color spread. Only stars brighter than $J = 16.5$ mag were used so as to minimize the inclusion of stars with large photometric 
errors. In order to treat differential reddening, a model grid was constructed from several isochrones reddened by $E(J-H) = 0.25, 0.64, 0.96, 1.27$, 
and 1.91 mag [equivalently $E(B-V) = 0.8, 2.0, 3.0, 4.0, 6.0$ mag],respectively. 

We attempted to estimate the masses of the PMS members, especially those in the embedded sub-clusters, using the model grid and 
compared the masses with those obtained from the optical data. We found the difference between the mass inferred from the 
near-IR CMD and that from the HRD to be a strong function of the age of the adopted isochrones, due to the luminosity evolution 
of PMS stars. For a given PMS star, the mass estimated from the reddened isochrone is larger for older isochrones. We found that 
the difference in masses from the two different methods showed a minimum for a grid of 0.1 Myr isochrones.

On the other hand, the mass of PMS stars ($J > 11.5$ mag) bluer than $J-H = 1.59$ was obtained from the the grid of isochrones 
with different ages (0.1, 0.5, 1, 2, 5, and 10 Myr) assuming the minimum reddening of $E(J-H) = 0.25$ mag, equivalent to $E(B-V) = 0.8$ mag. 
We compared the masses of all the PMS members with those from the optical data, and confirmed that there was a systematic difference of 
$-0.3$ to $0.4 M_{\odot}$ between them over the mass range of 0.2 to $1.6 M_{\odot}$. The systematic difference could be approximated by 
a combination of two straight lines. The mass from the near-IR data was converted to the mass scale obtained from the optical data using 
those relations. Finally, the masses of four early-type members and two members observed only in either $J$ or $H$ were obtained from 
analysis of the optical data. 

In order to examine the metallicity effect on the mass estimation, we also inferred the masses of the PMS stars 
using the PMS star evolution models for the half solar metallicity. The mass difference from the two models ($Z = Z_{\odot}$ 
and $Z=1/2Z_{\odot}$) is about 0.13$M_{\odot}$. It was confirmed that this difference yields a negligible effect on the resultant IMF in the mass 
range of 1 -- 100$M_{\odot}$.

We obtained the cluster mass of $169M_{\odot}$ from the sum of the masses of all the identified members. This is definitely a 
lower limit because a number of sub-solar mass stars below the completeness limit were not considered. IC 2162 seems to be the smallest SFR 
among the young open clusters studied by our research group, e.g., $> 510 M_{\odot}$ for NGC 1624 and NGC 1931 \citep{LSB15}, 
$> 576 M_{\odot}$ for NGC 2264, $> 1,300 M_{\odot}$ for NGC 1893 \citep{LSK14b}, $> 2,600 M_{\odot}$ for NGC 6231 \citep{SSB13}, 
$> 7,400 M_{\odot}$ for Westerlund 2 \citep{HPS15}, and $> 50,000 M_{\odot}$ for Westerlund 1 \citep{LCS13}.

The IMF is, in general, expressed by the following relation \citep{Sp55}:
\begin{equation}
\xi (\log m) \equiv {N \over \Delta\log m \cdot S}
\end{equation}

\noindent where $N$, $\Delta\log m$, and $S$ represent the number of stars in a given mass bin, the size of a logarithmic mass bin, and 
the area of the observed region, respectively. The size of the mass bin was set to be 0.4 to avoid uncertainties from small sample statistics. 
We assumed that the upper limit of stellar mass to be $100 M_{\odot}$, and adopted a larger bin size of 1 for the highest mass bin ($10 < M/M_{\odot} 
\le 100$) because the number of high-mass stars was insufficient to sample across various mass bins. We counted the number of stars 
in each mass bin, and then divided the total by the size of the mass bin and the area of IC 2162 (the area of our FOV). In order to prevent 
the binning effect, the IMF was re-derived for the same stars by shifting the mass bin by 0.2.

\begin{figure}[t!]
\centering
\includegraphics[angle=0,width=75mm]{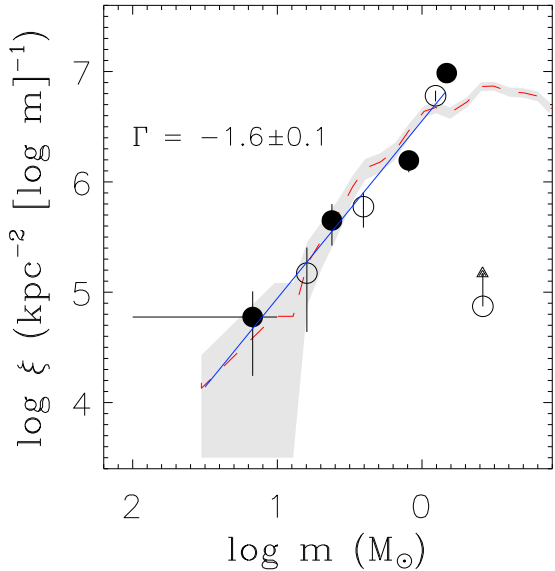}
\caption{The initial mass function (IMF) of IC 2162. The IMF (open and filled circles) was derived using different binning of the same stars to 
avoid the binning effect. The dashed line and shaded area represent the IMF of NGC 2264 and its uncertainty \citep{SB10}, respectively. 
The arrow indicates the the IMF below the completeness limit.}\label{fig9}
\vspace{5mm} %% add extra space ONLY when figures/tables are "colliding"!
\end{figure}

Figure~\ref{fig9} shows the IMF of IC 2162. We also plotted the IMF of the young open cluster NGC 2264 \citep{SB10} for comparison. 
The shapes of the IMFs are similar to each other in the mass range from $1 M_{\odot}$ to the upper limit of stellar mass. The IMF is generally 
quantified by its slope ($\Gamma$) to compare it with that of other SFRs. We computed the slope of the IMF using a linear least square 
fitting method. The slope of the IMF is about $\Gamma = -1.6 \pm 0.1$. This result is consistent with that of NGC 2264, although the slope 
is slightly steeper than that of the Salpeter/Kroupa IMF \citep{Sp55,K01}. It implies that the underlying star formation processes in IC 2162 
are optimized to produce low-mass stars rather than high-mass stars.

\section{SPECTRAL ENERGY DISTRIBUTION ANALYSIS}

The SED of the members was also analyzed using the SED fitting tool (called the SED fitter) of \citet{RWI07}. In 
order to increase the number of samples, we included 43 stars with X-ray and mid-IR excess emission in the membership catalogue. A total of 261 
members were used in this analysis. All the selected members were observed in more than 3 passbands, and their photometric errors were 
better than 0.15 mag in the optical to near-IR passbands and 0.20 mag in the mid-IR passbands.

In order to minimize the degrees of freedom, we limited the distance to $2.1 \pm 0.3$ kpc based on the result of the ZAMS fitting above. 
The total extinction $A_V$ was set to be 2.48 -- 100 mag. The SED fitter suggests various models with $\chi ^2$ values for a given SED. 
To select the most appropriate model among them, we followed the guidelines introduced by \citet{SB10}. The guidelines recommended 
the model that gives the smallest stellar mass out of the 10 suggested models if $\chi ^2 / \chi _{\mathrm{min}}  ^2 \leq 2.0$. 
If $\chi _{\mathrm{min}} ^2$ is less than 1.0, the mass-minimum model with $\chi ^2 < 2.0$ is adopted.  

The model adopted from the SED fitter provides various physical quantities of the disks and envelopes as well as stellar parameters (mass and age). 
The disk parameters such as mass, outer radius, and accretion rate evolve with time by a few orders of magnitude over 10 Myr. A strong 
correlation between the disk accretion rate and disk mass was also found [$\log(dM_{\mathrm{disk}}/dt) = -4.89 (\pm 0.12) + 1.15 (\pm 0.03) 
\log (M_{\mathrm{disk}}/M_{\star})$]. It appears that the envelope mass decreases dramatically after 1 Myr. While the angle of the cavity rises with time, 
its density, in contrast, declines as a function of time. The circumstellar extinction is proportional to the envelope mass and shows a sharp 
increase for stars with large envelope mass ($\log M_{\mathrm{env}}/M_{\star} > -3$). Stars with a massive envelope accrete much more 
material than low-mass counterparts. These correlations between the results obtained from the SED fitter are very similar to those found by 
\citet[see their Figure 3 and 4]{SB10}.

We also compared the stellar mass and age inferred from the SED fitter with those from the analysis of the HRD. For 
less reddened members, the masses obtained from the different methods are reasonably consistent, while the SED fitter 
underestimated the ages of the PMS members. On the other hand, the SED fitter tends to overestimate the masses and ages of 
highly reddened stars. This discrepancy has also been reported by \citet[see their Figure 2]{SB10}.

\section{Summary}
IC 2162 is an active SFR in which very young sub-clusters are embedded. We present a multiwavelength 
study of the SFR as part of the SOS project. This work provided homogeneous optical photometric data as well as a 
comprehensive result for the young stellar population in IC 2162.

The main ionizing sources in Sh 2-255 and 257 are two B0 stars (ALS 19 and HD 253327). A total of 218 
members were identified from the various photometric diagrams, the X-ray source list 
\citep{MPZ11}, and the YSO list \citep{CAH08}. It appears from the two early-type MS members that IC 2162 is reddened by at least 
$E(B-V) = 0.8$ mag. A large portion of the color spread in the near-IR CMD may be attributed to the presence of severe differential 
reddening, because a large number of stars are actually embedded in the molecular cloud behind the H{\scriptsize \textsc{II}} bubbles. The reddening law 
toward IC 2162 was investigated with various color excess ratios and the ratio of total-to-selective extinction found to
be the normal value ($R_V = 3.1$). It implies that the dust evolution in the front side of the SFR has been completed. 
We also revisited the distance to IC 2162 with the ZAMS fitting method and determined a distance of $2.1 \pm 0.3$ kpc. 

The ages of the members were estimated from the HRD using several evolutionary models \citep{SDF00,EGE12} for the solar metallicity. The median age 
of the optically visible stars in IC 2162 was about 1.3 Myr, and an age spread of 3.3 Myr was found. We derived the IMF of IC 2162 
by analyzing the HRD and the ($J, J-H$) diagram. The shape of the IMF is similar to that of the nearby young open cluster 
NGC 2264 in the mass range  $1 M_{\odot}$ to the upper limit of stellar mass. The slope of the IMF was $\Gamma = -1.6 \pm 0.1$, 
which is slightly steeper than that of Salpeter/Kroupa IMF. This result indicates that it was low-mass star formation that mostly took place 
throughout IC 2162. The lower limit on the cluster mass ($> 169 M_{\odot}$) was also constrained from the sum of the masses of all the identified members.

The SEDs of the members were also analyzed with the SED fitter \citep{RWI07}. The results included the disk and envelope parameters 
of PMS stars as well as stellar parameters such as age and mass. The properties of the disk and envelope were investigated as a function 
of stellar age or mass, respectively. The discrepancy between the results from the SED fitter and the analysis based on the 
HRD was also pointed out.

%%% ACKNOWLEDGMENTS (IF ANY) %%%%%%%%%%%%%%%%%%%%%%%%%%%%%%%%%%%%%%%%

\acknowledgments

This work was partly supported by a National Research Foundation of Korea grant funded 
by the Korean Government (Grant No. 20120005318) and partly supported by the Korea Astronomy 
and Space Science Institute (Grant No. 2015183014).

%%% CALL LIST OF REFERENCES (natbib STYLE) %%%%%%%%%%%%%%%%%%%%%%%%%%

\end{document}